\documentclass[preprintnumbers,amsmath,amssymb]{revtex4}
\usepackage{dcolumn}
\usepackage{bm}
\usepackage{graphicx,subfigure,xcolor}
\usepackage{braket}
\usepackage[normalem]{ulem}
\usepackage{balance}
\usepackage{comment}
\usepackage{xcolor}
\usepackage{amsmath}
\usepackage{amssymb}
\usepackage{eucal}
\usepackage{mathrsfs}
\usepackage{amsthm}
\usepackage{epstopdf}
\usepackage{textcomp}
\usepackage{braket}
\usepackage{pdfpages}
\usepackage[utf8]{inputenc}
\usepackage[T1]{fontenc}

\begin{document}

\title{Spontaneous emission of an atom near an oscillating mirror}

\author{Alessandro Ferreri $^{1}$\footnote{alessandro.ferreri@upb.de}}

\author{Michelangelo Domina$^{2}$\footnote{domina.michelangelo@gmail.com}}

\author{Lucia Rizzuto$^{2,3}$\footnote{lucia.rizzuto@unipa.it}}

\author{Roberto Passante$^{2,3}$\footnote{roberto.passante@unipa.it}}

\affiliation{$^1$ Department of Physics, Paderborn University, Warburger Strasse 100, D-33098 Paderborn, Germany}
\affiliation{$^2$ Dipartimento di Fisica e Chimica, Universit\`{a} degli Studi di Palermo, Via Archirafi 36, I-90123 Palermo, Italy}
\affiliation{$3$ INFN, Laboratori Nazionali del Sud, I-95123 Catania, Italy}

\begin{abstract}
We investigate the spontaneous emission of one atom placed near an oscillating reflecting plate.
We consider the atom modeled as a two-level system, interacting with the quantum electromagnetic field in the vacuum state, in the presence of the oscillating mirror.  We suppose that the plate oscillates adiabatically, so that the time-dependence of the  interaction Hamiltonian is entirely enclosed in the time-dependent mode functions, satisfying the boundary conditions at the plate surface, at any given time.
Using time-dependent perturbation theory, we  evaluate the transition rate to the ground-state of the atom, and show that it depends on the time-dependent atom-plate distance. We also show that the presence of the oscillating mirror significantly affects the physical features of the spontaneous emission of the atom, in particular the spectrum of the emitted radiation. Specifically, we find the appearance of two symmetric lateral peaks in the spectrum, not present in the case of a static mirror, due to the modulated environment. The two lateral peaks are separated from the central peak by the modulation frequency, and we discuss the possibility to observe them with actual experimental techniques of dynamical mirrors and atomic trapping.
Our results indicate that a dynamical (i.e. time-modulated) environment can give new possibilities to control and manipulate also other radiative processes of two or more atoms or molecules nearby, for example their cooperative decay  or the resonant energy transfer.
\end{abstract}

\maketitle

%\keyword{Spontaneous emission; Dynamical environments; Cavity Quantum Electrodynamics}

\newcommand{\wo}{\omega_0}
\newcommand{\wk}{{\omega_k}}
\newcommand{\wkp}{\omega_{k'}}
\newcommand{\ewo}{e^{i\omega_0 t}}
\newcommand{\ewom}{e^{-i\omega_0 t}}
\newcommand{\ewop}{e^{i\omega_0 t'}}
\newcommand{\ewomp}{e^{-i\omega_0 t'}}

\def\bk{{\bf k}}
\def\bkp{{\bf k}'}
\def\bkj{{\bf k}j}
\def\br{{\mathbf r}}
\def\bR{{\mathbf R}}
\def\wp{{\omega_p}}
\def\bfm{{\bf f}}

\newcommand{\bmu}{\boldsymbol\mu}
\newcommand{\vac}{\vert vac\rangle}
\newcommand{\ewk}{e^{i\omega_k t}}
\newcommand{\ewkm}{e^{-i\omega_k t}}
\newcommand{\ewkp}{e^{i\omega_k' t}}
\newcommand{\ewkmp}{e^{-i\omega_k' t}}
\newcommand{\ekr}{e^{i{\bf k}\cdot{\bf r}}}
\newcommand{\ekrm}{e^{-i{\bf k}\cdot{\bf r}}}
\newcommand{\ekri}{e^{i{\bf k}\cdot{\bf r}_i}}
\newcommand{\ekrmi}{e^{-i{\bf k}\cdot{\bf r}_i}}
\newcommand{\ekrA}{e^{i{\bf k}\cdot{\bf r}_A}}
\newcommand{\ekrmA}{e^{-i{\bf k}\cdot{\bf r}_A}}
\newcommand{\ekrAp}{e^{i{\bf k}'\cdot{\bf r}_A}}
\newcommand{\ekrmAp}{e^{-i{\bf k}'\cdot{\bf r}_A}}
\newcommand{\skj}{\sum_{{\bf k}j}}
\newcommand{\skjp}{\sum_{{\bf k'}j'}}
\newcommand{\akjd}{a^{\dag}_{{\bf k}j}}
\newcommand{\akj}{a_{{\bf k}j}}
\newcommand{\akjdp}{a^{\dag}_{{\bf k'}j'}}
\newcommand{\akjp}{a_{{\bf k'}j'}}
\newcommand{\ekj}{\hat{{\bf e}}_{{\bf k}j}}
\newcommand{\ekjp}{{\bf e}_{{\bf k'}j'}}

\section{\label{sec:1}Introduction}

Recent advances in quantum optics techniques and atomic physics have opened new perspectives for cavity quantum electrodynamics and solid state physics, making possible engineering systems with a tunable atom-photon coupling. Nowadays, the possibility to tailor and control radiative processes through suitable environments is of crucial importance in many different areas, ranging from condensed matter physics to quantum optics and quantum information theory \cite{raimond01,chang18}.

One of the most fundamental quantum processes is the spontaneous emission of radiation by atoms \cite{milonni94}. Purcell in $1946$ first suggested that spontaneous emission is not an unvarying property of the atoms, but it can be controlled (enhanced or inhibited) through the environment \cite{purcell46}. Physical properties of spontaneously emitted radiation depend strongly on the environment where the atom is placed: modifying the photon density of states and vacuum field fluctuations allows to change the spontaneous emission rate \cite{kleppner81,lodahl15}.
Many  physical systems have been explored in the literature to investigate this important process. These include, for example, atoms in cavities or waveguides \cite{kleppner81,hulet85,scully97,mok19,Gu17}, quantum dots in photonic crystals or in a medium with a photonic band gap \cite{john94,lambropoulos00,lodahl04}, and quantum emitters in metamaterials  \cite{newman13}.
Spontaneous decay of excited atoms in the presence of a driving laser field has been also investigated \cite{lewenstein87}.
Many experiments showing modifications of spontaneous emission of atoms in external environments (a single mirror, optical cavities, photonic crystals and waveguides, for example), have been also performed  \cite{drexhage74,goy83,heinzen87,eschner01,noda07}.

These investigations have shown how a structured environment, such as a cavity or a medium with periodic refractive index, can be exploited to control and tailor the spontaneous decay, as well as energy shifts of atomic levels, resonance and dispersion interactions between atoms, or the resonant energy transfer between atoms or molecules
\cite{shahmoon13,shahmoon14,incardone14,haakh15,notararigo18,fiscelli18}.

New interesting features appear when the boundary conditions on the field, or some relevant parameter of the system, change in time. Dynamical environments, whose optical properties change periodically in time, have been recently investigated, in particular in connection with the dynamical Casimir and Casimir-Polder effects \cite{dodonov10,crocce02,antezza14,souza19}. Dynamical Casimir effect has been first observed in superconducting circuit devices \cite{johansson09,wilson11}, that are also very promising devices for observing Unruh and Hawking effects, as well as other phenomena related to the quantum vacuum with time-dependent boundary conditions \cite{nation12}. The role of virtual photons exchange between moving mirrors in transferring mechanical energy between them has been recently investigated \cite{DiStefano19}. Also, time-dependent Casimir-Polder forces under non adiabatic conditions have been studied, showing that forces usually attractive may become repulsive under non equilibrium conditions \cite{shresta03,vasile08,messina10,haakh14,armata16}.

The spontaneous emission rate and the emission spectrum of an atom inside a dynamical (time-modulated)
photonic crystal, when its transition frequency is close to the gap of the crystal, have been recently investigated by the authors, finding modifications strictly related to the time-dependent photonic density of states \cite{calajo17}.
These findings suggested that a dynamical environment can give further possibilities to control radiative processes of atoms, which is of fundamental importance for many processes in quantum optics and its applications.
In this framework, the main aim of the present paper is to investigate the effects of a different kind of dynamical (time-dependent) environment, specifically an oscillating mirror, on the spontaneous decay of one atom in the vacuum, discussing both the decay rate and the emitted spectrum. As discussed later in this paper, this system appears within reach of actual experimental techniques of atomic trapping and dynamical mirrors.
Spontaneous emission of a two-level atom near an oscillating plate has been recently investigated in \cite{glaetze10} using a simple model, where  the quantized electromagnetic field is modeled as two one-dimensional fields, and in the rotating wave approximation; in \cite{glaetze10} it was also assumed that only field modes propagating within a small solid angle toward the mirror, and reflected back onto the atom, are affected by the mirror oscillation, while all other modes are assumed unaffected by its motion. These assumptions were justified by considering a specific atom-mirror-detector experimental setup, and in the spectrum they evaluate the photon population only in directions perpendicular and parallel to the atom-mirror direction. In our paper we instead consider a more general model, where the complete (three-dimensional, with all modes) electromagnetic field is quantized with the time-dependent boundary conditions determined by the (adiabatically) oscillating mirror. The contribution of all modes, whichever their propagation direction with respect to the mirror, is thus included in our calculation of the decay rate and of the spectrum, and we take into account the influence of the mirror's motion on photons propagating in all directions; also, a general orientation of the atomic dipole moment is considered. We also show that our system and our model are feasible with current experimental techniques, allowing observation of the effects we predict.

As mentioned above, we consider the full three-dimensional quantum electromagnetic field in the presence of a reflecting wall that oscillates adiabatically, so that the time dependence of the Hamiltonian is entirely enclosed in the time-dependent mode functions, satisfying the boundary conditions at the oscillating plate at any given time.
By using time-dependent perturbation theory, we first evaluate the transition rate of the atom from the excited to the ground state, and show that, as a consequence of the motion of the conducting mirror, it depends on time. Because of our adiabatic approximation, its time dependence follows the law of motion of the plate. Moreover, we show that the oscillatory motion of the mirror significantly affects also other physical features of the spontaneous emission of the atom, in particular the spectrum of the emitted radiation. We find that, for times larger than the inverse of the mirror's oscillation frequency, two lateral peaks in the spectrum appear; their distance from the central peak is equal to the oscillation frequency of the mirror (smaller peaks at a distance twice the mirror's oscillation frequency are also present). These peaks, contrarily to the case investigated in \cite{calajo17} for an atom in a dynamical photonic crystal, are symmetric with respect to the central peak. All this allows a sort of fine tuning of the emitted radiation exploiting the environment, and it could be relevant when other resonant processes are considered, for example the cooperative spontaneous decay of two or more atoms, or the resonance energy transfer between two atoms or molecules.

The paper is organized as follows. In Section \ref{sec:2}, we introduce our system, and investigate the decay rate of the two-level system in the presence of the oscillating mirror. In Section \ref{sec:3} we investigate the spectrum of the radiation emitted by the atom, and discuss its main physical features. Section \ref{sec:4} is devoted to our conclusive remarks.

\section{\label{sec:2} Spontaneous emission rate of one atom near an oscillating mirror}

\begin{figure}[!htbp]
\centering\includegraphics[width=6.0 cm]{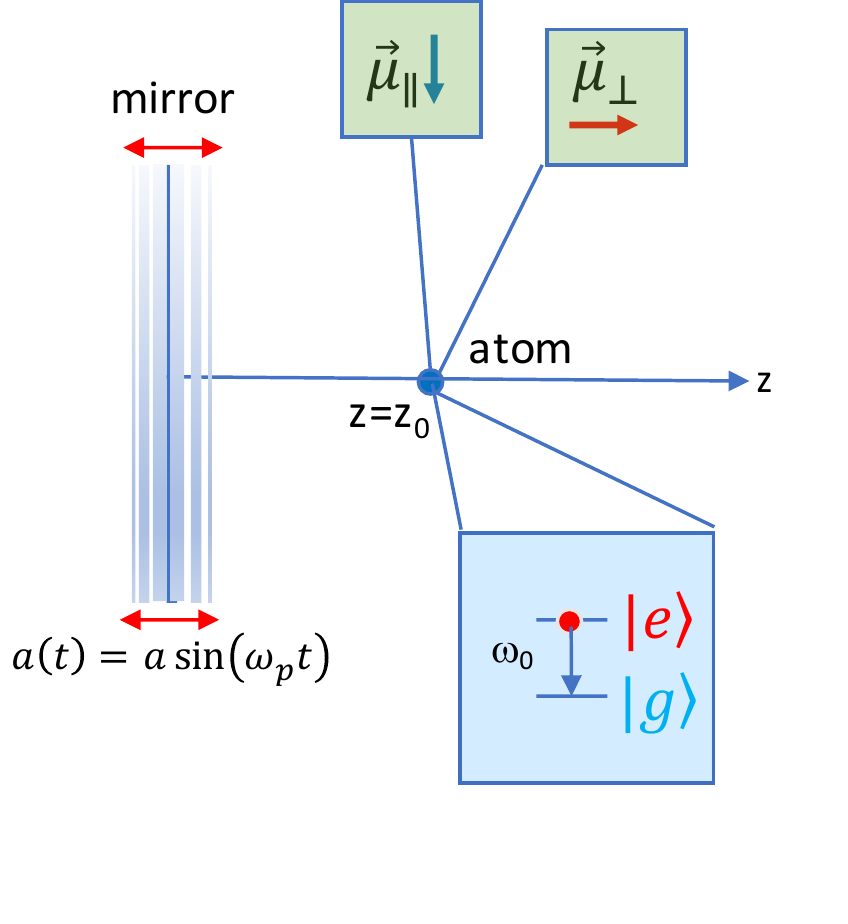}
\caption{Sketch of the system: one atom, modelled as a two-level system, is placed in front an oscillating mirror. The atomic dipole moment can be oriented parallel or perpendicular to the oscillating reflecting plane.}
\label{fig:1}
\end{figure}

Let us consider an atom, modeled as a two-level system with atomic transition frequency $\omega_0$, located in the half-space $z>0$ near an infinite perfectly conducting plate. Let us suppose that the mirror oscillates with a frequency $\omega_p$, along a prescribed trajectory
$a(t) = a \sin(\wp t)$, where $a$ is the oscillation amplitude of the plate, and $z=0$ is its average position. Although a mechanical motion with high oscillation frequencies is very difficult to obtain, it can be simulated by a dynamical mirror, that is a slab whose dielectric properties are periodically changed (from transparent to reflecting, for example), as obtained in proposed experiments for detecting the dynamical Casimir and Casimir-Polder effect \cite{braggio05,antezza14,agnesi09}. Dynamical mirrors have been recently obtained in the laboratory, with oscillation frequencies up to several GHz. They are based on a superconducting cavity with one wall covered by a specific semiconductor layer, having a high mobility of the carriers and very short recombination times. A train of laser pulses, with multigigahertz repetition rate, is then sent to the semiconductor layer: it creates a plasma sheet, that periodically changes the semiconductor layer from transparent to reflecting, thus simulating a mechanical motion of the cavity wall with frequencies that cannot be reached through a mechanical motion \cite{braggio05,agnesi09}. The atom can be kept at a fixed position by atomic trapping techniques \cite{reinhard08}.
Our physical system is pictured in Figure \ref{fig:1}, showing also relevant orientations, parallel and perpendicular, of the atomic transition dipole moment with respect to the plate.

We wish to investigate the effect of the mirror motion on the decay features (mainly decay rate and spectrum) of the two-level atom placed nearby, and interacting with the quantum electromagnetic field, initially in its vacuum state.
We assume that the reflecting plate oscillates adiabatically, and that its maximum velocity is such that $v_{p}=a\wp\ll c$, in order to have a nonrelativistic motion. Our adiabatic approximation is satisfied if the oscillation frequency of the plate $\omega_p$ is much smaller than the atomic transition frequency $\omega_0$,
and is also much smaller than the inverse of the time taken by a photon, emitted by the excited atom, to travel the atom-plate distance ($\omega_p\ll c/z_0$, where $z_0$ is the average atom-plate distance). In this case the atom instantaneously \textit{follows} the plate's motion. Real photons emission from the oscillating mirror by the dynamical Casimir effect can be thus neglected. These conditions are satisfied for typical values of the relevant parameters, currently achievable in the laboratory: $\wo\sim 10^{15}\,$s$^{-1}$, $\omega_p\sim 10^{9}\,$s$^{-1}$, and $z_0\sim 10^{-6} \,$m.
Under these assumptions, the mode functions of the field, satisfying the boundary conditions at the plate surface, depend explicitly on time. We may obtain their expression by generalizing the usual expressions of the field mode functions for a static mirror \cite{milonni94}, to the dynamical case using the instantaneous time-dependent atom-plate distance.
The mode functions can be written in the following form, after separating their vector part,
\begin{equation}
\label{eq:9}
[\bfm_{\bk j}(\br(t))]_\ell = [\ekj]_\ell f^{(\ell)}(\bk,\br(t)) ,
\end{equation}
where $\ekj$ are polarization unit vectors, $\bk$ is the wavevector, $j=1,2$ the polarization index, and $\ell =x,y,z$. The expression of the scalar functions $f^{(\ell )}(\bk,\br(t))$, that do not depend from the polarization $j$, is
\begin{eqnarray}
\label{eq:1}
f^{(x)}(\bk ,\br(t))&=& g_x(k_x,k_y)\sin\left[k_z z(t)\right] ,\\
\label{eq:2}
f^{(y)}(\bk ,\br(t))&=& g_y(k_x,k_y)\sin\left[k_z z(t)\right] ,\\
\label{eq:3}
f^{(z)}(\bk ,\br(t))&=& g_z(k_x,k_y)\cos\left[k_z z(t)\right] ,
\end{eqnarray}
where we have indicated with $g_\ell (k_x,k_y)$ ($\ell =x,y,z$) the time-independent part of the mode functions (\ref{eq:1}-\ref{eq:3}), given by
\begin{eqnarray}
\label{eq:1a}
g_x(k_x,k_y) &=& \sqrt{8}\cos\left[k_x\left(x+\frac L 2\right)\right]
\sin\left[k_y\left(y+\frac L 2\right)\right] ,\\
\label{eq:2a}
g_y(k_x,k_y) &=& \sqrt{8}\sin\left[k_x\left(x+\frac L 2\right)\right]
\cos\left[k_y\left(y+\frac L 2\right)\right] ,\\
\label{eq:3a}
g_z(k_x,k_y) &=& \sqrt{8}\sin\left[k_x\left(x+\frac L 2\right)\right]
\sin\left[k_y\left(y+\frac L 2\right)\right] .
\end{eqnarray}
In Eqs. (\ref{eq:1a}-\ref{eq:3a}), $L$ is the side of a cubic cavity of volume $V=L^3$ where the field is quantized (the cavity walls are at $x=\pm L/2$, $y=\pm L/2$, $z=0$, that is the average position of the oscillating mirror, and $z=L$); then, the limit $L \rightarrow \infty$ is taken, in order to recover the single oscillating mirror at $z=0$. Also, $z(t)$ is the time-dependent atom-plate distance, that changes in time according to the equation of motion $z(t) = z_0 - a(t) = z_0 - a \sin(\wp t)$.

The Hamiltonian of our system, in the Coulomb gauge and in the multipolar coupling scheme, within the dipole approximation \cite{compagno95,craig98,salam08,passante18}, is:
\begin{eqnarray}
\label{eq:4}
&\ & H=\hbar\wo S_z+\skj\hbar\wk \left(\akjd\akj+\frac 1 2 \right)+H_I ,
\end{eqnarray}
where $H_I$ is the interaction term, given by
\begin{equation}
\label{eq:5}
H_I = -i\sqrt{\frac{2 \pi c}{\hbar V}}\skj \sqrt{k}\left[ \bmu\cdot{\bf{f}}_{\bk j}({\br},t)\right]\left(S_+ + S_-\right)\left(\akj-\akjd \right).
\end{equation}
In (\ref{eq:5}), $\bmu$ is the matrix element of the atomic dipole moment operator between the ground and the excited states (assumed real), $S_z$ and $S_{\pm}$ the atomic pseudospin operators, $\akj$ ($\akjd$) the bosonic annihilation (creation) operators of the electromagnetic field. We note that, due to the adiabatic approximation, the interaction Hamiltonian is time-dependent through the mode functions only, while the field operators are the same as in the static case.

We now calculate the probability that the atom, initially excited with the field in the vacuum state, decays to its ground state at time $t$ by emitting one photon with wavevector ${\bk}$ and polarization $j$. Time-dependent perturbation theory up to the first order in the atom-field coupling gives
\begin{equation}
\label{eq:8}
\lvert c_{\bk j}(t)\rvert^2 = \frac{2\pi c k}{\hbar V}\int_0^{t}\int_0^{t}dt' dt''\left[\bmu\cdot\bfm_{\bk j}(\br(t'))\right]
\left[\bmu\cdot\bfm_{\bk j}(\br(t''))\right]e^{i(\wk-\wo)(t''-t')} .
\end{equation}

After polarization sum, using the relation,
\begin{equation}
\label{eq:10}
\sum_j(\ekj)_{\ell}(\ekj)_m = \delta_{\ell m}-\hat{k}_{\ell}\hat{k}_m ,
\end{equation}
with $\ell ,m=x,y,z$, Eq.(\ref{eq:8}) becomes
\begin{eqnarray}
\label{eq:11}
\lvert c_{\bk }(t)\rvert^2 &=& \sum_j \lvert c_{\bk j}(t)\rvert^2 = \frac{2\pi c k}{\hbar V}\sum_{\ell m}(\delta_{\ell m}-\hat{k}_{\ell}\hat{k}_m)\mu_\ell \mu_m \nonumber\\
&\ & \times  \int_0^{t}\! \int_0^{t} \! dt' dt'' f^{(\ell )}(\bk,\br(t')) f^{(m)}(\bk,\br(t'')) e^{i(\wk-\wo)(t''-t')} .
\end{eqnarray}

The expression (\ref{eq:11}) is valid for any orientation of the atomic dipole moment with respect to the plate.
In order to evaluate the decay probability, we sum (\ref{eq:11}) over $\bk$, obtaining $\lvert c(t) \rvert^2 = \sum_\bk \lvert c_\bk (t)\rvert^2$; then we take
continuum limit $V\rightarrow\infty$, $\sum_\bk \rightarrow V/(2\pi )^3 \int \! \! dk k^2 \int \! \! d\Omega_k$.

We now consider the specific cases of a dipole moment oriented parallel or orthogonal to the oscillating plate.
If the dipole moment is along the $x-$direction, then only the components $\ell=m=x$ are nonvanishing, and, after some algebra, we get
\begin{eqnarray}
\label{eq:12}
\lvert c(t)\rvert_x^2 &=& \frac {c \mu_x^2}{(2\pi)^2}
\int_{k_0-\Delta\omega/c}^{k_0+\Delta\omega/c} dk k^3\int d\Omega_k \left[1-\sin^2(\theta)\cos^2(\phi)\right] \nonumber \\
&\ & \times \int_0^{t} \!\!\int_0^{t} \!\!\! dt' dt'' f^{(x)}(\bk,\br(t'))f^{(x)}(\bk,\br(t''))e^{i(\wk-\wo)(t''-t')} ,
\end{eqnarray}
where we have limited the integration over $k$ to a band of width $\Delta \omega /c$ around $k_0=\omega_0/c$. This is justified by the fact that only (resonant) field modes with a frequency around the atomic transition frequency $\omega_0=ck_0$ give a relevant contribution to the integral over $k$. At the end of the calculation, we will take the limit $\Delta \omega \rightarrow \infty$.

Substituting the explicit expression of the scalar mode functions (\ref{eq:1}) into (\ref{eq:12}), after some algebra we find
\begin{eqnarray}
\label{eq:13}
\lvert c(t)\rvert^2_x &=& \frac{2\mu_x^2}{\hbar}k_0^3\Bigg[\frac{2}{3}t
- \int_0^t dt' \frac{\sin(2k_0z_0-2k_0a\sin(\omega_p t'))}{2k_0z_0-2k_0a\sin(\omega_pt')} \nonumber \\
&\ &-\int_0^t dt' \frac{\cos(2k_0z_0-2k_0a\sin(\omega_p t'))}{(2k_0z_0-2k_0a\sin(\omega_p t'))^2}
+\int_0^t dt' \frac{\sin(2k_0z_0-2k_0a\sin(\omega_p t'))}{(2k_0z_0-2k_0a\sin(\omega_p t'))^3}\Bigg] .
\end{eqnarray}
For a dipole moment oriented along the $y-$direction, we obtain for symmetry the same result (\ref{eq:13}), after substitution of $\mu_x$ with $\mu_y$.

In the case of a dipole moment orthogonal to the mirror, that is $\ell=m=z$, using the same procedure above, from (\ref{eq:11}) we obtain
\begin{equation}
\lvert c(t)\rvert^2_z = \frac {c \mu_z^2}{(2\pi)^2}\int_{k_0-\Delta\omega/c}^{k_0+\Delta\omega/c} dk k^3\int d\Omega_k[1-\cos^2 (\theta)]\int_0^{t}\int_0^{t}dt' dt'' f^{(z)}(\bk,\br(t')f^{(z)}(\bk,\br(t''))e^{i(\wk-\wo)(t''-t')} .
\label{eq:14}
\end{equation}
After some algebra, we get
\begin{equation}
\lvert c(t)\rvert^2_z = \frac{2\bmu_z^2}{\hbar}k_0^3\left[\frac{2}{3}t -\int_0^t \! \! dt' \frac{\cos(2k_0z_0-2k_0a\sin(\omega_p t'))}{(2k_0z_0-2k_0a\sin(\omega_p t'))^2} +\int_0^t \! \! dt' \frac{\sin(2k_0z_0-2k_0a\sin(\omega_p t'))}{(2k_0z_0-2k_0a\sin(\omega_p t'))^3}\right] .
\label{eq:15}
\end{equation}

We wish to stress that our results given by Equations (\ref{eq:13}) and (\ref{eq:15}) are valid within our adiabatic approximation.
meaning that the oscillation frequency of the plate is much smaller than both the atomic transition frequency ($\omega_p \ll \wo$) and the inverse of the time taken by a light signal to cover the atom-plate distance ($\omega_p \ll c/z_0$). Typical experimental values,
$\wo\sim 10^{15}\,$s$^{-1}$, $\omega_p\sim 10^{10}\,$s$^{-1}$ and $z_0 \sim 10^{-6} \,$m, well satisfy these conditions.

From (\ref{eq:13}) and (\ref{eq:15}) we can obtain the corresponding decay rates by taking their time derivative, $\Gamma_{x(y)} (z_0,t) = \frac d{dt}\lvert c(t)\rvert^2_{x(y)}$ for a dipole moment oriented parallel to the oscillating mirror (i.e. oriented along the $x-$ or $y-$direction), and $\Gamma_z (z_0,t) = \frac d{dt}\lvert c(t)\rvert^2_z$ for a dipole moment perpendicular to the plate. For a dipole moment randomly oriented, $\mu_x^2=\mu_y^2=\mu_z^2=\mu^2/3$, we finally get the (time-dependent) decay rate
\begin{eqnarray}
\label{eq:16}
\Gamma (z_0,t) &=& \Gamma_x (z_0,t) +\Gamma_y (z_0,t) +\Gamma_z (z_0,t)
= A_{12} \Bigg\{ 1 - \frac{\sin [2k_0(z_0-a\sin (\omega_p t))]}{2k_0(z_0-a\sin (\omega_p t))}
\nonumber \\
&\ &-2 \frac{\cos [2k_0(z_0-a\sin (\omega_p t))]}{[2k_0(z_0-a\sin (\omega_p t))]^2} +2
\frac{\sin [2k_0(z_0-a\sin (\omega_p t))]}{[2k_0(z_0-a\sin (\omega_p t))]^3}
\Bigg\} ,
\end{eqnarray}
where $A_{21}=4\mu^2k_0^3/3\hbar$ is the Einstein coefficient for spontaneous emission.
Our result (\ref{eq:16}) has a simple physical interpretation: it has the same structure of the rate for a static wall (see, for example Ref. \cite{milonni94}), but with the atom-wall distance replaced by the time-dependent distance $z_0-a\sin (\omega_p t)$, as indeed expected on a physical ground due to the adiabatic hypothesis.

For small oscillations of the  plate, keeping terms up to the first order in $a$, we obtain
\begin{eqnarray}
\label{eq:18}
\Gamma^{(1)}(z_0,t)
&\simeq& A_{21}\left[ 1-\frac{\sin(2k_0z_0)}{2k_0z_0}-2\frac{\cos(2k_0z_0)}{(2k_0z_0)^2}+ 2\frac{\sin(2k_0z_0)}{(2k_0z_0)^3} \right]\nonumber\\
&\ & + A_{21} \frac{a}{z_0}\sin(\omega_p t)\left[ \cos(2k_0 z_0)
-3\frac{\sin(2k_0z_0)}{2k_0z_0} - 6\frac{\cos(2k_0z_0)}{(2k_0z_0)^2}+6\frac{\sin(2k_0z_0)}{(2k_0z_0)^3}\right] .
\end{eqnarray}
Second and higher-order terms are negligible when $a/z_0 \ll 1$, and $k_0 z_0$ of the order of unity or less.
The expression above gives the total decay rate of our two-level atom near the oscillating mirror. When compared to the analogous quantity in the static case, the main difference is the presence of a time-dependent term. In particular, the quantity in the first line of (\ref{eq:18}) is the familiar decay rate of an atom near a static perfectly reflecting plate. Whereas, the other terms (second row of (\ref{eq:18})) depend on time, and describe the effect of the adiabatic motion of the conducting plate. They oscillate in time according to the oscillatory motion of the mirror, coherently with our adiabatic approximation. These new terms are of the order of $a/z_0$, and give a time and space  modulation of the decay rate directly related to the dynamics of the environment.

\section{\label{sec:3}Spectrum of the radiation emitted}

We now show that other relevant features appear in the radiation emitted by the atom in the presence of the (adiabatically) oscillating boundary, specifically significant changes of its spectrum.
We now evaluate the probability amplitude $c(\bk j,t)$ that the atom, initially prepared in the excited state with the field in the vacuum state, decays to its ground state by emitting a photon in the field mode $(\bk j)$. Time-dependent perturbation theory (at first order in the atom-field coupling) gives
\begin{eqnarray}
\label{eq:19}
c_{\bk j}(t) = \sqrt{\frac{2\pi c k}{\hbar V}} \int_0^t dt' \left(\bmu\cdot\bfm_{\bk j}(\br(t')\right)e^{i(\wk-\wo)t'} .
\end{eqnarray}

As in the previous section, we make the approximation of small oscillations of the plate ($a \ll z_0$), and expand the mode functions of the field, keeping only terms up to the second order in $a$. A straightforward calculation gives
\begin{eqnarray}
\label{eq:20}
\left[\bfm_{\bk j}(\br(t))\right]_{x(y)} &\simeq& [\ekj]_{x(y)} \, g_{x(y)}(k_x,k_y)\Big[ \sin(k_z z_0) -k_z a \cos(k_z z_0)\sin(\omega_p t)\nonumber \\
&\ & -\frac 12 (k_z a)^2 \sin(k_z z_0)\sin^2(\omega_p t)\Big] , \\
\label{eq:21}
\left[\bfm_{\bk j}(\br(t))\right]_{z} &\simeq& [\ekj]_z \, g_{z}(k_x,k_y)\Big[\cos(k_z z_0) +k_z a \sin(k_z z_0)\sin(\omega_p t)\nonumber \\
&\ & -\frac 12 (k_z a)^2 \cos(k_z z_0)\sin^2(\omega_p t)\Big] ,
\end{eqnarray}
where  the functions $g_i(k_x,k_y)$ ($i=x,y,z$) have been defined in Eqs. (\ref{eq:1a}-\ref{eq:3a}).
In the vector notation, indicating with $\br_0=(0,0,z_0)$ the atom's position, the instantaneous atom-mirror distance is $\br (t) = \br_0 - {\bf a} \sin (\omega_pt)$, with ${\bf a}=(0,0,a)$, and we have
\begin{equation}
\label{eq:20a}
\bfm_{\bk j} (\br(t)) = \bfm_{\bk j}^{(0)}(\br_0) -a \bfm_{\bk j}^{(1)}(\br_0) \sin (\omega_p t) + \frac 12 a^2 \bfm_{\bk j}^{(2)}(\br_0) \sin^2(\omega_p t) + \ldots ,
\end{equation}
where $\bfm_{\bk j}^{(0)}(\br_0)=\bfm_{\bk j}(\br_0)$ are the mode functions for a static mirror at $z=0$ \cite{milonni94,power82}, evaluated at $\br_0=(0,0,z_0)$ (see Eqs. (\ref{eq:9}-\ref{eq:3})),
and $\bfm_{\bk j}^{(1)}(\br_0)$ and $\bfm_{\bk j}^{(2)}(\br_0)$ come from the first- and second-order corrections in the expansion (\ref{eq:20a}) of $\bfm_{\bk j}(\br(t))$ in powers of $a$,
\begin{equation}
\label{eq:26}
[\bfm_{\bk j}^{(1)}(\br_0)]_i = \frac{\partial}{\partial z_0}[\bfm_{\bk j}(\br_0)]_i ,
\end{equation}
\begin{equation}
\label{eq:26a}
[\bfm^{(2)}_{\bk j}(\br_0)]_i = \frac{\partial^2}{\partial z_0^2}[\bfm_{\bk j}(\br_0)]_i = - k_z^2 [\bfm^{(0)}_{\bk j}(\br_0)]_i .
\end{equation}
with $i=x,y,z$ (see Eqs. (\ref{eq:9}-\ref{eq:3})).
Putting the expansion (\ref{eq:20a}) into (\ref{eq:19}), taking into account (\ref{eq:26}) and (\ref{eq:26a}), and integrating over time, we obtain
\begin{eqnarray}
\label{eq:22}
c_{\bk j}(t)&\simeq& \sqrt{\frac{2\pi c k}{\hbar V}}\left\{\bmu\cdot \bfm_{\bk j}^{(0)}(\br_0)\frac{e^{i(\wk-\wo)t}-1}{i(\wk-\wo)} \right. \nonumber \\
&\ &  - \frac a2 \bmu\cdot
\bfm_{\bk j}^{(1)}(\br_0)\left[\frac{1-e^{i(\wk-\wo+\omega_p)t}}{\wk-\wo+\omega_p}-\frac{1-e^{i(\wk-\wo-\omega_p)t}}{\wk-\wo-\omega_p}\right]
- \frac 14 a^2 k_z^2 \bmu\cdot \bfm_{\bk j}^{(0)}(\br_0)
\nonumber \\
&\ & \ \left. \times \left[ \frac{e^{i(\wk-\wo)t}-1}{i(\wk-\wo)}
+ \frac{1-e^{i(\wk-\wo +2\omega_p)t}}{2i(\wk-\wo +2\omega_p)} + \frac{1-e^{i(\wk-\wo -2\omega_p)t}}{2i(\wk-\wo -2\omega_p)} \right] \right\} .
\end{eqnarray}

By taking the squared  modulus of Eq. (\ref{eq:22}), we now evaluate the transition probability at time $t$. After some algebra, and keeping only terms up to second order, we find
\begin{equation}
\label{eq:27}
P_{\bk j}(\omega_p,t) = \lvert c_{\bk j}(t)\rvert^2 \simeq P_{\bk j}^{(0)}(t) + P_{\bk j}^{(dyn)}(\omega_p,t) ,
\end{equation}
where
\begin{equation}
\label{eq:28}
P_{\bk j}^{(0)}(t)=\frac{2\pi c k}{\hbar V} [\bmu\cdot \bfm_{\bk j}^{(0)}(\br_0) ]^2\frac{\sin^2[(\wk -\wo) t/2]}{[(\wk -\wo/2)]^2}
\end{equation}
is the $0-$th order contribution, coinciding with that obtained for a static boundary, while $P_{\bk j}^{(dyn)}(t)$ is the change to the spectrum due to the motion of the plate. The dynamical correcting term is obtained as
\begin{eqnarray}
\label{eq:29}
P_{\bk j}^{(dyn)}(\omega_p,t) &=&
-\frac{2\pi c ka}{\hbar V} [\bmu\cdot \bfm_{\bk j}^{(0)}(\br_0)]  [\bmu\cdot \bfm_{\bk j}^{(1)}(\br_0) ]  h_1(\wk -\wo,\omega_p,t)
\nonumber \\
&\ &+\frac{\pi c k a^2}{2\hbar V}[\bmu\cdot \bfm_{\bk j}^{(1)}(\br_0) ]^2 h_2(\wk -\wo,\omega_p,t)
\nonumber \\
&\ & -\frac{\pi c k a^2k_z^2}{\hbar V}[\bmu\cdot \bfm_{\bk j}^{(0)}(\br_0) ]^2 h_3(\wk -\wo,\omega_p,t) ,
\end{eqnarray}
where we have defined the functions
\begin{eqnarray}
\label{eq:32}
h_1(\wk -\wo,\omega_p,t)&=&\sin(\omega_p t/2)\frac{\sin[(\wk -\wo )t/2]}{(\wk -\wo )/2} \nonumber \\
&\ & \times \left(\frac{\sin[(\wk -\wo +\omega_p)t/2]}{(\wk -\wo +\omega_p)/2}
+\frac{\sin[(\wk -\wo -\omega_p)t/2]}{(\wk -\wo -\omega_p)/2}\right) ,
\end{eqnarray}
\begin{eqnarray}
\label{eq:33}
h_2(\wk -\wo,\omega_p,t)&=& \frac{\sin^2[(\wk -\wo +\omega_p)t/2])}{(\wk -\wo +\omega_p)^2/4}
+\frac{\sin^2[(\wk -\wo-\omega_p)t/2]}{(\wk -\wo -\omega_p)^2/4}  \nonumber \\
&\ & -2\cos (\omega_P t) \frac {\sin[(\wk -\wo +\omega_p)t/2] \sin [(\wk -\wo -\omega_p)t/2)]}{(\wk -\wo +\omega_p)(\wk -\wo -\omega_p)/4} ,
\end{eqnarray}
and
\begin{eqnarray}
\label{eq:33a}
h_3(\wk -\wo,\omega_p,t)&=& \frac{\sin^2[(\wk -\wo ) t/2]}{[(\wk -\wo )/2]^2} -\cos (\omega_pt) \frac {\sin [(\wk -\wo) t/2]}{\wk -\wo}
\nonumber \\
&\ & \times \left( \frac {\sin [(\wk -\wo +2\omega_p)t/2]}{\wk -\wo +2\omega_p} + \frac {\sin [(\wk -\wo -2\omega_p)t/2]}{\wk -\wo -2\omega_p} \right)
\end{eqnarray}

We stress that our results above include the effect of the mirror's motion on photons propagating in any direction. Inspection of Eqs. (\ref{eq:28}) and (\ref{eq:29}), taking into account (\ref{eq:32}), (\ref{eq:33}) and (\ref{eq:33a}), clearly shows the modifications of the spectrum of the spontaneously emitted radiation: the presence of two lateral peaks, at frequencies $\wk=\wo\pm\omega_p$, in addition to the ordinary resonance peak at $\wk=\wo$. From (\ref{eq:33a}), the presence of (smaller) lateral peaks at the frequencies $\wk=\wo\pm 2\omega_p$ is also evident, indicating a sort of nonlinear behavior of the system.

In order to obtain an explicit expression of the emitted spectrum  as  a function of the frequency, we sum over polarizations and perform the angular integration, obtaining the probability density for unit frequency. Keeping only terms up to second order in $a$ ($a < z_0, \, k_0^{-1})$, a lengthy but straightforward calculation gives
\begin{equation}
\label{eq:34}
P_\wk (\omega_p,t) =\frac V{(2\pi )^3} \frac {\wk^2}{c^2} \sum_j \int \! \! d\Omega P_{\bk j}(\omega_p,t)
= P_\wk^{(0)} (t) + P_\wk^{(dyn)} (\omega_p,t) ,
\end{equation}
with
\begin{equation}
\label{eq:36}
P_\wk^{(0)} (t)=\frac{1}{2\pi} A_{21}\frac{\sin^2[(\wk -\wo )t/2]}{[(\wk -\wo )/2]^2}\left[1-\frac{\sin(2k_0z_0)}{2k_0 z_0}-2\frac{\cos(2k_0z_0)}{(2k_0 z_0)^2} +2\frac{\sin(2k_0z_0)}{(2k_0 z_0)^3}\right] ,
\end{equation}
and
\begin{eqnarray}
\label{eq:37}
P_\wk^{(dyn)} (\omega_p,t)&=&\frac{A_{21}}{2\pi} \Big\{ \frac{a}{2z_0}h_1(\wk -\wo ,\omega_p,t)\left[\cos(2k_0 z_0)-3\frac{\sin(2k_0z_0)}{2k_0 z_0}-6\frac{\cos(2k_0z_0)}{(2k_0 z_0)^2} +6\frac{\sin(2k_0z_0)}{(2k_0 z_0)^3}\right] \nonumber \\
& & + \left(\frac{ak_0}2 \right)^2 h_2(\wk -\wo ,\omega_p,t)\left[\frac 1 3 + \frac{\sin(2k_0z_0)}{2k_0 z_0}+4\frac{\cos(2k_0z_0)}{(2k_0 z_0)^2}-12\frac{\sin(2k_0z_0)}{(2k_0 z_0)^3} \right.
\nonumber\\
&\ &\left. - 24\frac{\cos(2k_0z_0)}{(2k_0 z_0)^4}+24\frac{\sin(2k_0z_0)}{(2k_0 z_0)^5}\right]
\nonumber \\
&\ & +\frac {(ak_0)^2}2  h_3(\wk -\wo ,\omega_p,t)\left[-\frac 13 +\frac{\sin(2k_0z_0)}{2k_0 z_0}+4\frac{\cos(2k_0z_0)}{(2k_0 z_0)^2} -12\frac{\sin(2k_0z_0)}{(2k_0 z_0)^3} \right.
\nonumber \\
&\ &\left. - 24\frac{\cos(2k_0z_0)}{(2k_0 z_0)^4} +24\frac{\sin(2k_0z_0)}{(2k_0 z_0)^5}\right] \Big\}.
\end{eqnarray}

Thus, in the dynamical case, we find, apart the usual peak at $\wk=\wo$, the presence of two lateral peaks of the radiation emitted at frequencies $\wk=\wo\pm\omega_p$, related to the presence of the energy denominators in Eq. (\ref{eq:37}) (see also (\ref{eq:32}) and (\ref{eq:33})). Other, smaller, lateral peaks at $\wk=\wo\pm 2\omega_p$ are also present, given by the last term in (\ref{eq:37}).

Our results above, being based on a second-order perturbative expansion on the oscillation amplitude $a$ of the mirror, are valid for small oscillations, $a/z_0 \ll 1$, $(ak_0)^2 \ll 1$, and for $k_0z_0$ of the order of unity or less. These conditions are feasible for typical experimental values. For example, if $\wo = ck_0 \sim 10^{15} \,$s$^{-1}$, we can reasonably take  $z_0 \sim 10^{-6} \,$m and $a \sim 10^{-7} \,$m. These values are also fully compatible with our adiabatic assumption, using realistic values of $\omega_p$ of a few GHz. Higher oscillation amplitudes can be exploited in the case of Rydberg states, which have much lower values of $k_0$ \cite{Gallagher88}; in such a case, the oscillation frequency of the plate $\omega_p$ must be smaller accordingly, due to our adiabatic approximation.

\begin{figure}[!htbp]
\centering\includegraphics[width=8.5cm]{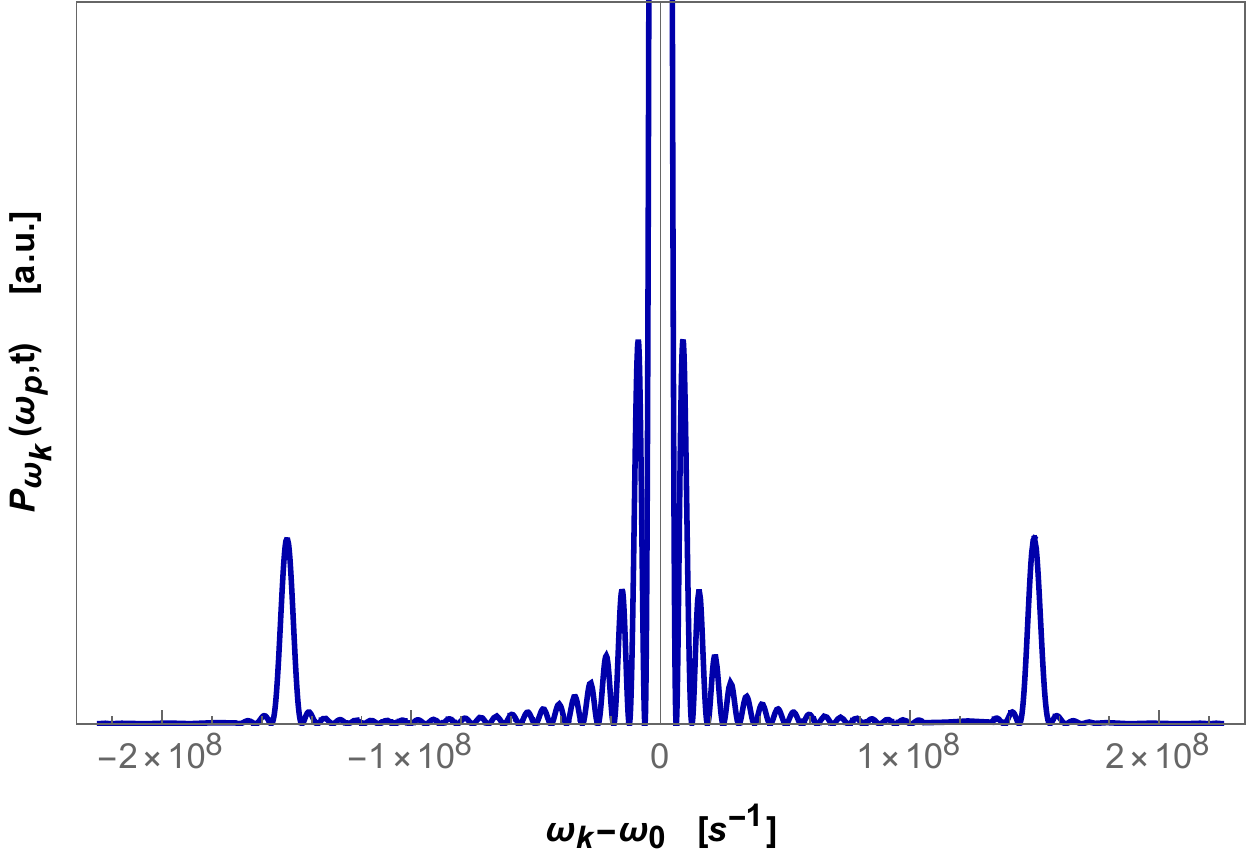}
\caption{Plot of the spectrum $P_\wk (\omega_p,t)$ of emitted radiation from an atom near an oscillating mirror, in arbitrary units, in terms of $\wk -\wo$. The two lateral peaks in the photon spectrum are symmetric, and  separated from the atomic transition frequency by the oscillation frequency of the plate (in the figure, $\omega_p = 1.5 \cdot 10^8 \,$s$^{-1}$; the other numerical values of the relevant parameters are: $\wo  = 10^{15} \,$s$^{-1}$, $z_0 = 10^{-6} \,$m; $a = 2 \cdot 10^{-7} \,$m; $t = 10^{-6} \,$s).}
\label{fig:2}
\end{figure}

Figure \ref{fig:2} shows a plot of the emitted spectrum by the excited atom in the limit of long times ($t\gg \omega_p^{-1}$, but small enough to make valid our perturbative approach), as a function of $\wk -\wo$, showing the two lateral peaks at $\omega = \wo \pm \omega_p$. We found a similar behavior of the spectrum for a two-level atom located inside a dynamical photonic crystal \cite{calajo17}; in that case, however, the two lateral peaks were strongly asymmetric due to the different density of states at the edges of the the photonic band gap. Instead, in the present case of an atom in the vacuum space near an oscillating boundary, the two lateral peaks are symmetric, because the photonic density of states is essentially the same at the peaks' frequencies (the rapid oscillations in the figure, as well as in the next Figure \ref{fig:3}, come from the fact that we are considering finite times; thus, the physical meaning should be extracted from the envelope of the curve plotted).

Inspection of (\ref{eq:37}) shows that the lateral peaks become more and more evident for times larger than $\omega_p^{-1}$. Figure \ref{fig:3} gives the spectral density as a function of time, for $\omega_p = 1.5 \cdot 10^9 \,$s$^{-1}$, clearly showing the lateral peaks growing with time, and becoming sharper and well identifiable when $t \gg 2\pi \omega_p^{-1}$.

\begin{figure}[!htbp]
\centering\includegraphics[width=11cm]{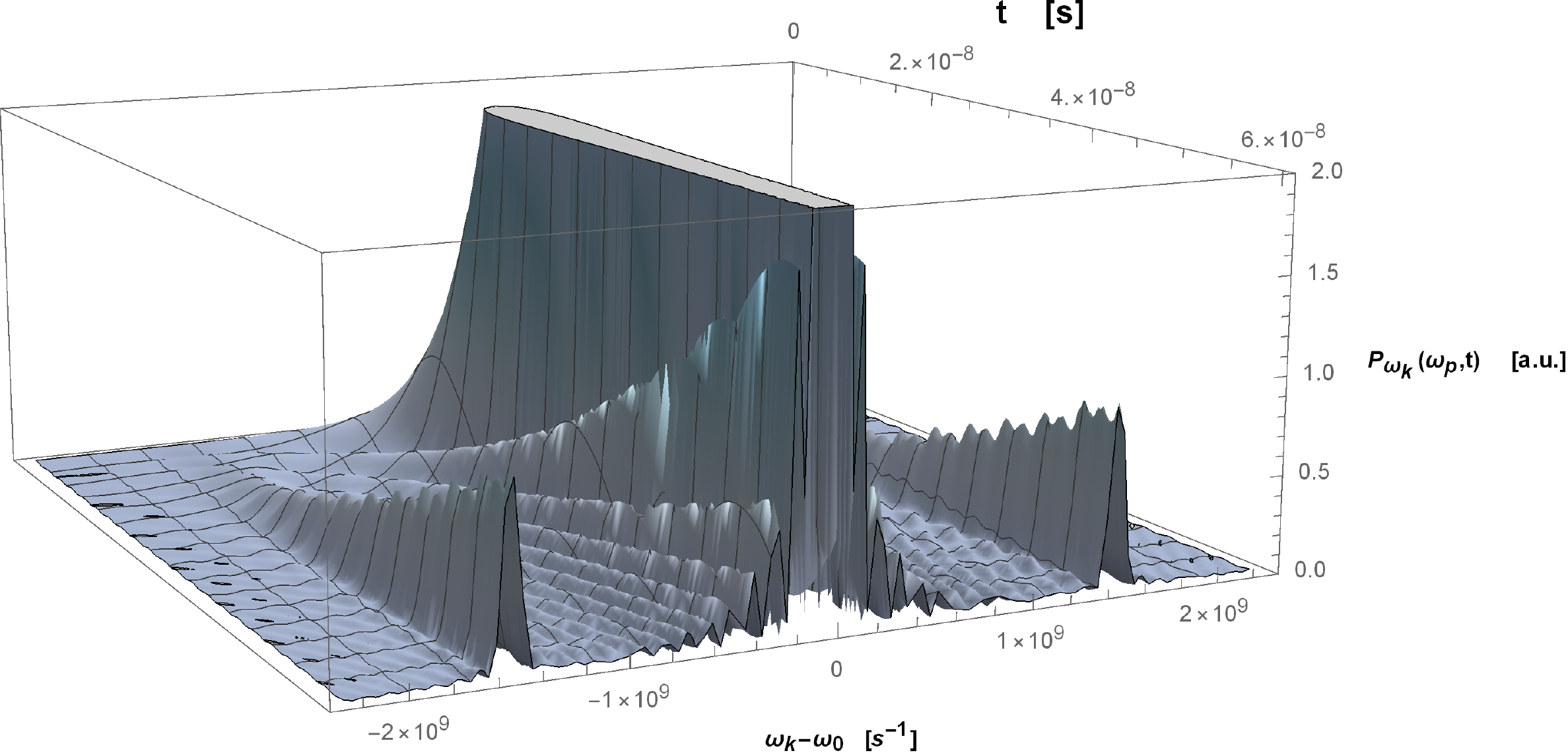}
\caption{Spectral density of the emitted radiation, in arbitrary units, in terms of $\wk -\wo$ and time $t$, with $\omega_p = 1.5 \cdot 10^9 \,$s$^{-1}$. The figure shows that the lateral peaks become more and more evident as time goes on, specifically when $t \gg 2\pi \omega_p^{-1}$. The other parameters are: $\wo  = 10^{15} \,$s$^{-1}$, $z_0 = 10^{-6}$m; $a = 2 \cdot 10^{-7} \,$m.}
\label{fig:3}
\end{figure}

In order to resolve the lateral peaks in the emitted spectrum, their distance $\omega_p$ from the central peak must be larger that the natural linewidth of the emission line. For example, if we consider the optical transition between the levels $n=3$ and $n=2$ of the hydrogen atom, the natural width of the line is $ \sim 10^8 \,$s$^{-1}$, and an oscillation frequency of $\nu_p = \omega_p/2\pi \sim 10^9 \,$s$^{-1}$ or more is thus sufficient to resolve the lateral lines; such a frequency can be actually reached with the technique of dynamical mirrors \cite{braggio05,agnesi09} mentioned in the Introduction.
In the case of Rydberg atoms, the plate oscillation frequency must be much smaller; however, also the natural linewidth of the transition can be very small if the Rydberg atoms are prepared in a circular state \cite{Gallagher88}. Also, actual experimental techniques make feasible trapping of low-density gases of Rydberg atoms with micrometric precision, and for sufficiently long times \cite{pillet09,antezza14,Barredo19}, as well as trapping atoms at submicrometric distances from a surface \cite{Thompson13}. This should make easier to experimentally observe the effects found in the system here investigated, in particular the lateral peaks of the spectrum, rather than in the case of atoms in a dynamical photonic crystal previously investigates \cite{calajo17}.

Our results show that spontaneous emission can be controlled (enhanced or suppressed) by  modulating in time the position of a perfectly reflecting plate, and that the spectrum of the emitted radiation can be controlled through the oscillation frequency of the plate. This suggest the possibility to control also other radiative processes through modulated (time-dependent) environments, for example the cooperative decay of two or more atoms, or the resonance energy transfer between atoms or molecules. These systems will be the subject of a future publication.

\section{\label{sec:4}Conclusions}
In this paper, we have investigated the features of the spontaneous emission rate and of the emitted spectrum of one atom, modeled as a two level system, near an oscillating perfectly reflecting plate, in the adiabatic regime. We have discussed in detail the effect of  the motion of the mirror on the spontaneous decay rate, and shown that it is modulated in time.
We have also found striking modifications of the emission spectrum, that exhibits, apart the usual peak at $\omega=\wo$,  two new lateral peaks separated from the atomic transition frequency by the oscillation frequency of the plate. The possibility to observe these lateral peaks with current experimental techniques of dynamical mirrors and atomic trapping has been also discussed. Our findings for the spontaneous emission indicate that modulated environments can be exploited to manipulate and tailor the spontaneous emission process; also, they strongly indicate that a dynamical environment could be successfully exploited to modify, activate or inhibit also other radiative processes of atoms or molecules nearby.

\vspace{6pt}

%\authorcontributions{All authors equally contributed to this paper.}

\begin{acknowledgments}
The authors gratefully acknowledge financial support from the Julian Schwinger Foundation and MIUR.
\end{acknowledgments}

%\conflictsofinterest{The authors declare no conflict of interest.}

\end{document}